\documentclass[twocolumn,showpacs,prc]{revtex4}
\usepackage{epsf} 
\usepackage{dcolumn}
\usepackage{bm}

\begin{document}

\title{Absorption of Nuclear Gamma Radiation by Heavy Electrons  \\ 
on Metallic Hydride Surfaces}

\author{A. Widom}
\affiliation{Physics Department, Northeastern University, 110 Forsyth Street,
Boston MA 02115}
\author{L. Larsen}
\affiliation{Lattice Energy LLC, 175 North Harbor Drive, Chicago IL 60601}

\begin{abstract}
Low energy nuclear reactions in the neighborhood of metallic hydride 
surfaces may be induced by ultra-low momentum neutrons. Heavy electrons 
are absorbed by protons or deuterons producing ultra low momentum neutrons 
and neutrinos. The required electron mass renormalization is provided by the 
interaction between surface electron plasma oscillations and surface proton 
oscillations. The resulting neutron catalyzed low energy nuclear reactions 
emit copious prompt gamma radiation. The heavy electrons which induce the 
initially produced neutrons also strongly absorb the prompt 
nuclear gamma radiation, re-emitting soft photons. Nuclear hard photon 
radiation away from the metallic hydride surfaces is thereby strongly 
suppressed.  
\end{abstract}

\pacs{24.60.-k, 23.20.Nx}
\maketitle

\section{Introduction \label{Intro}}

Low energy nuclear reactions (LENR) may take place in 
the neighborhood of metallic hydride 
surfaces\cite{Iwamura:2002,Focardi:1998}.
The combined action of surface electron density plasma oscillations and 
surface proton oscillations allow for the production of heavy mass 
renormalized electrons. A heavy electron, here 
denoted by \begin{math} \tilde{e}^-  \end{math}, may produce 
ultra-low momentum neutrons via the reaction\cite{Marchak:1969}  
\begin{equation}
\tilde{e}^{-}+p^+\to n+\nu_e.
\label{intro1}
\end{equation}
Once the ultra low momentum  neutrons are created, other more complex 
low energy nuclear reactions may be catalyzed\cite{WidomLarsen:2005}. 
Typically, neutron catalyzed nuclear reactions release energy in large part
by the emission of prompt hard gamma radiation. However, the copious 
gamma radiation and neutrons have not been observed away from the 
metallic hydride surface. Our purpose is to theoretically explain this 
experimental state of affairs. In particular, we wish to explore the theoretical 
reasons why copious prompt hard gamma radiation has {\em not} been 
observed for LENR on metallic hydride surfaces. 
The experimental fact that a {\em known} product particle is not observed 
far from the metallic hydride surface is related to the fact that the mean 
free path of the product particle to be converted to other particles is 
short. As an example of such arguments, we review in  Sec. \ref{NMFP},  why 
the mean free path of an ultra low momentum neutron is so short.
A short mean free path implies that the product particle never 
appears very far from the surface in which it was first created.  

For hard gamma radiation, the mean free path computation in metals is 
well known\cite{Heitler:1953,PDG:2000}. For normal metals, there 
exists {\em three} processes determining prompt gamma  photon mean 
free path. The processes are as follows:
\par \noindent 
(1) {\em The photoelectric effect:} The hard photon blasts a bound core 
electron out of the atom.
\par \noindent
(2) {\em Compton scattering from normal conduction electrons:}  The hard 
photon scatters off a very slowly moving conduction electron giving up a 
finite fraction of its energy to this electron. 
\begin{equation}
\gamma + e_i^-  \to  \gamma^\prime +e_f ^-
\label{normal1}
\end{equation}
The final photon \begin{math} \gamma^\prime   \end{math} is nonetheless fairly hard. 
\par \noindent 
(3) {\em Creation of electron-positron pairs:} The hard gamma photon creates 
an electron-positron pair. Kinematics disallows this one photon process in the 
vacuum. Pair production can take place in a metal wherein other charged particles 
can recoil during the pair production process. 
Roughly, the resulting mean free paths of hard prompt gamma photons 
is of the order of centimeters when all of the above above mechanisms 
are taken into account.  

Let us now consider the situation in the presence of heavy electrons. 
The processes are as follows:
\par \noindent  
(1) {\em The absence of a heavy electron photoelectric effect:} The surface 
heavy electrons are all conduction electrons. They do not occupy bound core 
states since the energy is much too high.  Thus, there should be no heavy 
electron photoelectric effect.  There will be an anomalously high surface 
electrical conductivity due to these heavy conduction electrons. 
This anomaly occurs as the threshold proton (or deuteron) density  for 
neutron catalyzed LENR is approached. 
\par \noindent
(2) {\em Compton scattering from heavy conduction electrons:} When the hard 
gamma photon is scattered from a heavy electron, the final state of the radiation 
field consists of very many very soft photons; i.e. the Compton Eq.(\ref{normal1}) is 
replaced by 
\begin{equation}
\gamma + \tilde{e}_i^-  \to  \sum \gamma_{soft} +\tilde{e}_f ^- .
\label{heavy1}
\end{equation}
It is the final state soft radiation, shed from the mass renormalized electrons, that 
is a signature for the heavy electron Compton scattering.  
\par \noindent 
(3) {\em Creation of heavy electron-hole pairs:} In order to achieve heavy electron 
pair energies of several MeV, it is not required to reach way down into the 
vacuum Dirac electron sea\cite{Dirac :1934}. The energy differences between 
electron states in the heavy electron conduction states is sufficient to pick up 
the ``particle-hole'' energies of the order of MeV. 
Such particle-hole pair production in conduction states of metals 
is in conventional condensed matter physics described by electrical conductivity.

In Sec. \ref{OPMFP}, the theory of electromagnetic propagation in metals 
is explored. We show that an optical photon within a metal has a very short 
mean free path for absorption. The short mean free path makes metals opaque to  
optical photons. The effect can be understood on the basis of the Bohr 
energy rule 
\begin{equation}
\Delta E=\hbar \omega .
\label{intro2}
\end{equation}
For optical frequencies, \begin{math}  \hbar \omega  \end{math}
is of the order of a few electron volts and typical particle-hole 
pair creation energies near the Fermi surface are also of the order 
of a few electron volts. The resulting strong electronic absorption 
of optical photons is most easily described by the metallic electrical 
conductivity. For hard photons with an energy of the order of a few MeV, 
there are ordinarily no electronic particle-hole solid state excitations 
with an energy spread which is so very large. A normal metal is thus 
ordinarily transparent to hard gamma rays. 

On the other hand, the non-equilibrium neutron catalysis of LENR 
near metallic hydride surfaces is due to heavy electrons with a 
renormalized mass having energy spreads of 
several MeV. These large energy spreads for the heavy surface electrons 
yield the mechanism for hard gamma ray absorption. This mechanism is 
explained in detail in Secs. \ref{PHE} and \ref{HGRA}. In 
Sec. \ref{PHEA}, we review the reasons for which a single electron in 
the vacuum {\em cannot} absorb a hard single massless photon. In 
Sec. \ref{PHEB}, we review the reasons why a single electron within a 
plain wave radiation beam {\em can} absorb a hard single massless photon. 
The proof requires the well established exact solutions of the Dirac 
wave functions in a plane wave radiation field\cite{Berestetskii:1997}
and contains also the proof of the induced electron mass renormalization 
in this radiation field. The electrical conductivity of induced 
non-equilibrium heavy electrons on the metallic hydride surface as 
seen by hard photons will explored in Sec. \ref{HGRA}. The energy spread of heavy 
electron-hole pair excitations implies that a high conductivity near 
the surface can persist well into the MeV photon energy range strongly 
absorbing prompt gamma radiation. An absorbed {\em hard} gamma photon 
can be re-emitted as a very large number of {\em soft} photons, e.g. infrared 
and/or X-ray. The  mean free path of a hard gamma photon estimated from  
physical kinetics\cite{Landau:1995} has the form 
\begin{equation}
L_\gamma \approx \left(\frac{3}{\pi}\right)^{1/3}
\left(\frac{1}{4\alpha }\right)
\frac{1}{\tilde{n}^{2/3}\tilde{l}}
\approx \frac{33.7}{\tilde{n}^{2/3}\tilde{l}}\ ,
\label{intro3}
\end{equation}
where \begin{math} \tilde{n}  \end{math} is the number of heavy electrons 
per unit volume, \begin{math}  \tilde{l}   \end{math} is the mean free path 
of a heavy electron and the quantum electrodynamic coupling strength is 
\begin{equation}
\alpha =\frac{e^2R_{vac}}{4\pi \hbar}=\frac{e^2}{\hbar c}
\approx \frac{1}{137.036} \ .
\label{intro4}
\end{equation}
The number density of heavy electrons on a metallic hydride surface 
is of the order of the number density of surface hydrogen atoms
when there is a proton or deuteron flux moving through the surface and 
LENR are being neutron catalyzed. These added heavy electrons produce 
an anomalously high surface electrical conductivity at the LENR threshold.  
Roughly, \begin{math} \tilde{n}^{2/3}\sim 10^{15}/{\rm cm}^2  \end{math}, 
\begin{math} \tilde{l}\sim 10^{-6}{\rm \ cm}  \end{math} so that the mean free 
path of a hard prompt gamma ray is 
\begin{math} L_\gamma \sim 3.4\times 10^{-8}\ {\rm cm} \end{math}.
Thus, prompt hard gamma photons get absorbed within less than a 
nanometer from the place wherein they were first created. The energy 
spread of the excited particle hole pair will have a cutoff of about 
\begin{math}  10\ {\rm MeV}  \end{math} based on the mass 
renormalization of the original electron. The excited heavy electron 
hole pair will annihilate, producing very many soft photons based on 
the photon spectrum which produced such a mass renormalization.
The dual role of the heavy electrons is discussed in 
the concluding Sec. \ref{CON}.  In detail, the heavy electrons are 
absorbed by protons creating ultra-low momentum neutrons and 
neutrinos which catalyze further LENR, e.g.  
subsequent neutron captures on nearby nuclei. The heavy electrons 
also allow for the strong absorption of prompt hard gamma radiation 
produced from LENR. 

\section{Neutron Mean Free Path \label{NMFP}}

Suppose there are \begin{math} n \end{math} neutron absorbers per unit 
volume with an absorption cross section \begin{math}  \Sigma  \end{math}. 
The mean free path \begin{math}  \Lambda  \end{math} of the neutron is 
then given by 
\begin{equation}
\Lambda ^{-1}=n\Sigma =\frac{4\pi \hbar n}{p}\Im m\ {\cal F}(0)
=\frac{4\pi \hbar nb}{p}\ ,
\label{NMFP1}
\end{equation}
where \begin{math} p \end{math} is the  neutron momentum and  
\begin{math}  {\cal F}(0)  \end{math} is the forward scattering 
amplitude. The imaginary part of the scattering length is denoted 
by \begin{math} b \end{math}. In terms of the ultra low momentum 
neutron wave length 
\begin{math} \lambda =(2\pi \hbar/p) \end{math}, 
Eq.(\ref{NMFP1}) implies  
\begin{equation}
\Lambda=\frac{1}{2n\lambda b}\ .
\label{NMFP2}
\end{equation}
The ultra low momentum neutron is created when a heavy electron is absorbed 
by one of many protons participating in a collective surface oscillation. 
The neutron wave length is thus comparable to the spatial size  of the 
collective oscillation, say 
\begin{math} \lambda \sim 10^{-3}\ {\rm cm} \end{math}. With (for example) 
\begin{math} b \sim 10^{-13}\ {\rm cm}   \end{math} and 
\begin{math} n\sim 10^{22}\ {\rm cm}^{-3}  \end{math}, one finds a neutron 
mean free path of \begin{math} \Lambda \sim 10^{-6}\ {\rm cm} \end{math}. 
An ultra low momentum neutron is thus absorbed within about ten nanometers 
from where it was first created. The likelihood that ultra low momentum 
neutrons will escape capture and thermalize via phonon interactions is very 
small. 

\section{Optical Photon Mean Free Path \label{OPMFP}}

Consider an optical photon propagating within a metal with 
conductivity \begin{math} \sigma  \end{math}. From Maxwell's 
equations 
\begin{eqnarray}
curl{\bf E} &=& -\frac{1}{c}\frac{\partial {\bf B}}{\partial t}, 
\nonumber \\
div{\bf B}&=&0,
\nonumber \\
curl{\bf B} &=& \frac{1}{c}
\left\{\frac{\partial {\bf E}}{\partial t}+4\pi {\bf J}\right\},
\label{OPMFP1}
\end{eqnarray} 
it follows that 
\begin{equation}
\left\{\frac{1}{c^2}\left(\frac{\partial}{\partial t}\right)^2 
-\Delta \right\}{\bf B}-\frac{4\pi }{c}curl{\bf J}=0.
\label{OPMFP2}
\end{equation}
Employing Ohms law in the form 
\begin{eqnarray} 
{\bf J}  &=& \sigma {\bf E}
\nonumber  \\   
curl\  {\bf J} &=&
-\frac{\sigma}{c}\left(\frac{\partial {\bf B}}{\partial t}\right), 
\label{OPMFP3}
\end{eqnarray}
yields the wave equation with dissipative damping  
\begin{equation}
\left\{\left(\frac{\partial}{\partial t}\right)^2 
+4\pi \sigma \left(\frac{\partial}{\partial t}\right)
-c^2\Delta \right\}{\bf B}=0.
\label{OPMFP4}
\end{equation}
The transition rate per unit time for the optical photon absorption 
is then \begin{math} 4\pi \sigma \end{math}. This argument yields an 
optical photon mean free path \begin{math} L  \end{math} given by
\begin{equation}
\frac{1}{L}=\frac{4\pi \sigma}{c}=R_{vac}\sigma 
\label{OPMFP5}
\end{equation}
wherein \begin{math} R_{vac} \end{math} is the vacuum impedance. In 
SI units, the optical photon mean free path is given by  
\begin{equation}
L=\frac{1}{R_{vac}\sigma }\ \ \ 
{\rm where}\ \ \ 
\frac{R_{vac}}{4\pi }\equiv 29.9792458\ {\rm Ohm}.  
\label{OPMFP6}
\end{equation}
For a metal with low resistivity 
\begin{equation}
\sigma ^{-1} \lesssim 10^{-5}\ {\rm Ohm\ cm},
\label{OPMFP7}
\end{equation}
the mean free path length of an optical photon obeys 
\begin{equation}
L \lesssim 3\times 10^{-8}\  {\rm cm}.
\label{OPMFP8}
\end{equation}
An optical photon in a metal is absorbed in less than 
a nanometer away from the spot in which it was born.
Thus, normal metals are opaque to visible light.

To see what is involved from a microscopic viewpoint, let us 
suppose an independent electron model in which occupation 
numbers are in thermal equilibrium with a Fermi distribution 
\begin{equation}
f(E)=\frac{1}{e^{(E-\mu)/k_BT}+1}\ .
\label{OPMFP9}
\end{equation}
If the conductivity is described in terms of elastic electron 
scattering from impurities or phonons, then the conductivity in 
a volume \begin{math} \Omega \end{math} containing conduction electrons 
is described by the Kubo formula\cite{Kubo:1957}
\begin{eqnarray}
& &{\Re e}\{ \sigma (\omega +i0^+)\} = -\frac{\pi }{6}
\left(\frac{e^2}{\Omega }\right) \times 
\nonumber \\ 
& &\sum_{i,f}\left[\ \frac{f(E_f)-f(E_i)}{E_f-E_i} \right]  
|{\bf v}_{fi}|^2
[\delta (\omega -\omega_{fi})  +  \delta (\omega +\omega_{fi}) ],
\nonumber \\  
& &\hbar \omega_{fi} = E_f-E_i,
\label{OPMFP10}
\end{eqnarray}
wherein \begin{math} {\bf v}_{fi}  \end{math} is a matrix element of 
the electron velocity operator \begin{math} {\bf v}  \end{math}. If one 
starts from the interaction between an electron and a photon in the form 
\begin{equation}
H_{int}=-\frac{e}{c}{\bf A\cdot v},
\label{OPMFP11}
\end{equation}
applies Fermi's Golden rule for photon absorption, averages over initial 
states and sums over final states, then the result for the frequency 
dependent optical photon mean free path \begin{math} L(\omega ) \end{math} 
is 
\begin{equation}
\frac{1}{L(\omega )}=\frac{4\pi }{c} {\Re e}\{ \sigma (\omega +i0^+)\} 
=R_{vac}  {\Re e}\{ \sigma (\omega +i0^+)\} ,
\label{OPMFP12}
\end{equation}
where Eq.(\ref{OPMFP10}) has been taken into account. Eqs.(\ref{OPMFP10}) 
and (\ref{OPMFP12}) are merely the microscopic version  of Eq.(\ref{OPMFP5}) 
which followed directly from Maxwell's equations and Ohm's law.
In thermal equilibrium , the energy differences between electron states are 
of the order of electron volts. As the photon frequency is increased to the 
nuclear physics scale of MeV, the electrical conductivity 
\begin{math}   {\Re e}\{ \sigma (\omega +i0^+)\}    \end{math} 
rapidly approaches zero. Thus, a metal in thermal equilibrium is 
almost transparent to hard nuclear gamma radiation. As will be discussed in 
what follows, for the surfaces of metallic hydrides in non-equilibrium 
situations with heavy electrons, strong absorption of nuclear gamma 
radiation can occur.  

\section{Hard Photons - Heavy Electrons \label{PHE}}

Heavy electrons appear on the surface of a metallic hydride in 
non-equilibrium situations. Sufficient conditions include 
(i) intense LASER radiation incident on a suitably rough metallic 
hydride surface, (ii) high chemical potential differences across the 
surface due electrolytic voltage gradients and (iii) high chemical potential 
differences across the surface due to pressure gradients. Under such 
non-equilibrium conditions, weakly coupled surface plasmon polariton 
oscillations and proton oscillations induce an oscillating electromagnetic field 
\begin{equation}
F_{\mu \nu }(x)=\partial_\mu A_\nu (x)-\partial_\nu A_\mu (x)
\label{PHE1}
\end{equation} 
felt by surface electrons. From a classical viewpoint, the electrons 
obey the Lorentz force equation 
\begin{equation}
\frac{d^2x^\mu}{d\tau ^2}=\frac{e}{c}F^\mu _{\ \ \nu }(x)
\frac{dx^\nu}{d\tau } .
\label{PHE2}
\end{equation}
From the quantum mechanical viewpoint, the electron wave function obeys 
the Dirac equation
\begin{equation}
\left\{\gamma^\mu \left(-i\hbar \partial_\mu 
-\frac{e}{c}A_\mu (x)\right)+mc\right\}\psi (x)=0.
\label{PHE3}
\end{equation}
The quantum motions are intimately related to the classical motions as can 
be seen by formulating the problem in terms of the Hamilton-Jacobi action 
\begin{math} S(x) \end{math}. To describe the classical orbits according 
to Eq.(\ref{PHE2}), one seeks a classical velocity field 
\begin{math} v^\mu (x) \end{math} obeying the 
Hamilton-Jacobi equations\cite{Landau:1975}
\begin{eqnarray}
mv_\mu (x) &=& \partial_\mu S(x)-\frac{e}{c}A_\mu (x),
\nonumber \\ 
v^\mu (x)v_\mu (x) &=& -c^2. 
\label{PHE4}
\end{eqnarray}
The orbits implicit in the second order Eqs.(\ref{PHE2}) may now be obtained 
by solving the first order equations of motion 
\begin{equation}
\frac{dx^\mu }{d\tau }=v^\mu (x).
\label{PHE6}
\end{equation}
From the quantum mechanical biewpoint, one seeks a solution to the Dirac 
Eq.(\ref{PHE4}) having the form 
\begin{equation}
\psi (x)=u(x)e^{iS(x)/\hbar }.
\label{PHE7}
\end{equation}
The classical velocity field \begin{math} v^\mu (x) \end{math} makes its 
appearance in the equation of motion for the spinor 
\begin{math} u(x) \end{math}; It is exactly 
\begin{equation}
\left\{\gamma^\mu \big(-i\hbar \partial_\mu 
+mv^\mu (x)\big)+mc\right\}u(x)=0.
\label{PHE8}
\end{equation}
Eqs.(\ref{PHE4}), (\ref{PHE7}) and (\ref{PHE8}) constitute the 
``unperturbed'' electron states in the classical electromagnetic 
field \begin{math} F_{\mu \nu } \end{math} describing {\em soft} 
radiation from a non-perturbation theory viewpoint. The {\em hard} gamma 
photons may {\em thereafter} be treated employing low order perturbation 
theory. Two specific examples should suffice to illustrate the point. 

\subsection{Free Electrons in the Vacuum \label{PHEA}}

A classical free electron has a Hamilton-Jacobi action which obeys 
\begin{eqnarray}
\partial_\mu S(x)\partial^\mu S(x) &+& m^2c^2 = 0,
\nonumber \\ 
S(x) &=& p_\mu x^\mu ,
\nonumber \\ 
p_\mu p^\mu  &=& -m^2c^2.
\label{PHEA1}
\end{eqnarray}
For a classical free electron, the velocity field is uniform in space 
and time; i.e.
\begin{eqnarray}
\frac{dx^\mu }{d\tau } &=& v^\mu =\frac{p^\mu }{m}\ ,
\nonumber \\  
x^\mu  &=& \left(\frac{p^\mu }{m}\right)\tau .
\label{PHEA2}
\end{eqnarray}
The free particle quantum theory solutions follow from Eqs.(\ref{PHE7}), 
(\ref{PHE8}), (\ref{PHEA1}) and (\ref{PHEA2}) according to  
\begin{eqnarray}
\psi (x) &=& u(p)e^{ip\cdot x/\hbar},
\nonumber \\  
\big(\gamma^\mu p_\mu &+& mc\big)u(p) = 0.
\label{PHEA3}
\end{eqnarray}
Eq.(\ref{PHEA3}) serves as the starting point for the computation 
of a single hard photon absorption (with wave vector 
\begin{math}  k  \end{math} and polarization 
\begin{math}  \epsilon \end{math}) by an electron in the vacuum. 
The {\em vanishing} amplitude is computed as 
\begin{eqnarray}
{\cal F}(e^-_i+\gamma \to e^-_f )  &=& 
\frac{i}{\hbar c^2}\int J^\mu _{fi}(x)A_\mu (x)d^4 x,  
\nonumber \\ 
{\cal F}(e^-_i+\gamma \to e^-_f )  &=& \frac{ie}{\hbar c}
\int \bar{\psi }_f\ (x) \gamma^\mu \psi_i (x)A_\mu (x)d^4 x,
\nonumber \\ 
{\cal F}(e^-_i+\gamma \to e^-_f )  &=&  
\left(\frac{ieA_\gamma }{\hbar c}\right) 
\epsilon_\mu \bar{u}(p_f)\gamma^\mu u(p_i) \times 
\nonumber \\ 
& &\int e^{i(p_i+\hbar k -p_f)\cdot x/\hbar}d^4x,
\nonumber \\ 
{\cal F}(e^-_i+\gamma \to e^-_f )  &=&  
\left(\frac{ieA_\gamma }{\hbar c}\right) 
\epsilon_\mu \bar{u}(p_f)\gamma^\mu u(p_i) \times 
\nonumber \\ 
& &(2\pi \hbar)^4\delta (p_i+\hbar k-p_f),
\nonumber \\ 
{\cal F}(e^-_i+\gamma \to e^-_f )  &=&  0.
\label{PHRA4}
\end{eqnarray} 
Of central importance is the impossibility of satisfying the 
four momentum conservation law 
\begin{math} p_i+\hbar k=p_f   \end{math}  for fixed electron mass 
\begin{math} p_i^2=p_f^2=-m^2c^2  \end{math} and zero 
photon mass, 
\begin{equation}
k^\mu k_\mu =0.
\label{PHRA5}
\end{equation}
Thus, hard photon absorption by a single electron in the vacuum is 
not possible. 

\subsection{Electrons and Electromagnetic Oscillations \label{PHEB}}

Suppose that the electron is in the field of {\em soft} radiation. For 
example, the electron may be within a plane wave laser radiation beam. 
Further suppose an additional {\em hard} gamma photon is incident upon 
the electron. In this case (unlike the vacuum case) the hard photon can 
indeed be absorbed. For a plane wave electromagnetic oscillation of the 
form 
\begin{eqnarray}
A_{soft}^\mu ( x) &=& a^\mu (\phi ),
\nonumber \\
\phi  &=& q_\mu x^\mu \ \ \ {\rm where}\ \ \ q^\mu q_\mu =0, 
\nonumber \\ 
\partial_\mu A_{soft}^\mu ( x) &=& 
q_\mu \frac{da^\mu (\phi)}{d\phi }=0,
\label{PHEB1}
\end{eqnarray}
the action function is\cite{Lifshitz:1975}
\begin{equation}
S(x)=p_\mu x^\mu + W_p(\phi ).
\label{PHEB2}
\end{equation}
To find the function \begin{math} W_p(\phi )  \end{math},
one may compute the velocity field
\begin{equation}
mv^\mu (x) = p^\mu -\frac{e}{c}a^\mu (\phi )+q^\mu W_p ^\prime (\phi ),
\label{PHEB3}
\end{equation}
and impose the Hamilton-Jacobi condition 
\begin{math} v^\mu v_\mu =-c^2  \end{math}. 
Solving this condition for \begin{math} W_p (\phi ) \end{math} 
taking Eq.(\ref{PHEB1}) into account yields 
\begin{eqnarray}
2(p\cdot q )W_p ^\prime (\phi) &=& \frac{2ep\cdot a(\phi )}{c}
-\left\{m^2 c^2 +p^2 +\frac{e^2a^2(\phi )}{c^2} \right\},
\nonumber \\ 
W_p (\phi) &=& \int_0^\phi W_p ^\prime (\tilde{\phi })d\tilde{\phi }. 
\label{PHEB4}
\end{eqnarray}
If \begin{math} W^\prime _p(\phi ) \end{math} remaines finite during a oscillation 
period, then on phase averaging  
\begin{math} \overline{W_p ^\prime }=\tilde{p}-p \end{math}. This leads to mass 
renormalization \begin{math} m\to \tilde{m} \end{math} of the electron in the 
laser field 
\begin{eqnarray}
\tilde{p}^2+(\tilde{m} c)^2 &=& 0,
\nonumber \\ 
(\tilde{m}c)^2 &=& (mc)^2+\left(\frac{e}{c}\right)\overline{a^2(\phi )}.
\label{PHEB5}
\end{eqnarray}
Importantly, the same electromagnetic oscillations which increase the 
electron mass, also allow for the absorption of hard gamma photons by 
the surface heavy electron. The gamma ray absorption amplitude for a 
heavy electron has the form\cite{Lifshitz :1997}  
\begin{eqnarray}
\mathcal{F}(\tilde{e}^-_i+\gamma \to \tilde{e}^-_f ) &=& 
\frac{i}{\hbar c^2}\int J^\mu _{fi}(x)A_\mu (x)d^4 x,
\nonumber \\ 
\mathcal{F}(\tilde{e}^-_i+\gamma \to \tilde{e}^-_f ) &=&
\left(\frac{ieA_\gamma }{\hbar c}\right) \epsilon_\mu \times 
\nonumber \\  
& & \int  \bar{u}_f\gamma u_i e^{i(\tilde{p}_i+\hbar k -\tilde{p}_f)\cdot x/\hbar}\times 
\nonumber \\ 
& & e^{i(\tilde{W}_i(\phi )-\tilde{W}_f(\phi ))/\hbar}d^4x,
\nonumber \\ 
\mathcal{F}(\tilde{e}^-_i+\gamma \to \tilde{e}^-_f ) &=&
\sum_n \epsilon_\mu \mathcal{F}_{i\to f}^\mu (n) \times
\nonumber \\ 
& &\delta (\tilde{p}_i+\hbar k-\tilde{p}_f-\hbar nq).
\label{PHEB6}
\end{eqnarray}
The conservation of four momentum for heavy electrons  
\begin{math} \tilde{p}_i+\hbar k=\tilde{p}_f+n\hbar q \end{math} as in 
Eq.(\ref{PHEB6}) differs the conservation of four momentum 
\begin{math} p_i+\hbar k=p_f \end{math} for the vacuum case in 
Eq.(\ref{PHRA4}). Only the heavy electrons can absorb a hard photon with 
four momentum \begin{math} \hbar k \end{math} emitting 
\begin{math} n \end{math} photons each with four momentum 
\begin{math} \hbar q \end{math}.
The transition rate per unit time per unit volume for the reaction 
\begin{equation}
\tilde{e}^-_i+\gamma \to \tilde{e}^-_f+n\gamma_{soft}
\label{PHEB7}
\end{equation}
is given by 
\begin{eqnarray}
\nu_{i\to f}(n) &=& \frac{(2\pi )^4}{c}
\left|\epsilon_\mu \mathcal{F}_{i\to f}^\mu (n)\right|^2\times 
\nonumber \\ 
& & \delta \left(\frac{\tilde{p}_f-\tilde{p}_i}{\hbar }+nq-k\right),
\nonumber \\ 
\tilde{p}_i^2 &=& \tilde{p}_f^2=-(\tilde{m}c)^2 ,
\nonumber \\ 
\nu_{i\to f} &=& \sum_n \nu_{i\to f}(n).
\label{PHEB8}
\end{eqnarray}
with a renormalized electron mass given in Eq.(\ref{PHEB5}).
Thus, for the exactly soluble plane wave low frequency oscillation, 
we have shown how a high frequency hard photon cam scatter 
off a heavy electron producing many low frequency soft photons. 

\section{Hard Photon  Absorption \label{HGRA}}

Let us now return to the heavy electron oscillations on the surface of 
metallic hydrides. The same heavy electrons required to produce neutrons 
for catalyzed LENR are also capable of absorbing prompt gamma 
radiation via the electrical conductivity which now extends to frequencies 
as high as about 
\begin{math} \hbar \omega_{max} \sim 10\  {\rm MeV} \end{math}. Just as 
for the optical photon case of Sec. \ref{OPMFP}, we may proceed via 
Maxwell's equations for the hard photon field
\begin{eqnarray}
\delta F_{\mu \nu} &=& \partial_\mu \delta A_\nu -\partial_\nu \delta A_\mu , 
\nonumber \\ 
\partial_\mu \delta A^\nu  &=& 0, 
\nonumber \\ 
\partial_\mu \delta F^{\mu \nu } &=& -R_{vac} \delta J^\mu.
\label{HBRA1}
\end{eqnarray}
The heavy electron current response to the prompt hard photon may be written as 
\begin{equation}
R_{vac} \delta J^\mu (x)=\int \Pi^{\mu \nu}(x,y;A)\delta A_\nu (y)d^4y.
\label{HBRA2}
\end{equation} 
Especially note that the heavy electron current response function 
\begin{math} \Pi  \end{math} depends on the soft radiation field which 
renormalized the electron mass in the first place. To lowest order in the 
quantum electrodynamic coupling in Eq.(\ref{intro4}),  we have the 
independent electron model relativistic Kubo formula in the one loop 
insertion  form\cite{Schwinger:1951}
\begin{eqnarray}
\left\{-i\gamma^\mu\left(\partial_\mu 
-i \frac{eA_\mu (x)}{\hbar c}\right) 
+\frac{mc}{\hbar }\right\}\mathcal{G}(x,y,;A) = \delta(x-y),
\nonumber \\ 
\Pi^{\mu \nu} (x,y;A) =
4\pi i\alpha \ tr
\left\{\gamma^\mu \mathcal{G}(x,y;A)
\gamma^\nu \mathcal{G}(y,x;A)\right\} . \  \
\label{HBRA3}
\end{eqnarray}
The equation of motion for the hard prompt gamma photon is thereby 
\begin{equation}
-(\partial_\lambda \partial^\lambda)\delta A^\nu (x)
-\int \Pi^{\mu \nu}(x,y;A)\delta A_\nu (y) d^4y = 0,
\label{HBRA4}
\end{equation}
wherein the damping is contained in \begin{math} \Pi  \end{math} 
which in turn depends on background electromagnetic field oscillations.

The {\em physical kinetics} estimate\cite{Ziman:1995} of 
the hard prompt gamma photon absorption 
is as follows: (i) In the energy regime of the heavy electron mass renormalization, 
the conductivity obeys  
\begin{eqnarray}
 \sigma_\gamma &\approx & \frac{1}{(3\pi^2)^{1/3}}
\left(\frac {e^2}{\hbar}\right)(\tilde{n}^{2/3}\tilde{l}) ,
\nonumber \\ 
{\rm for\ energies} &\ & 0.5\ {\rm MeV} \lesssim 
\hbar \omega_\gamma  \lesssim  10\ {\rm MeV},
\label{HBRA5}
\end{eqnarray} 
where the heavy electrons have a density per unit volume of 
\begin{math} \tilde{n} \end{math} and an electronic mean free path  
of \begin{math} \tilde{l} \end{math}. (ii) We estimate for typical 
metallic hydride LENR, the values 
\begin{equation}
\tilde{n}^{2/3}\sim 10^{15}/{\rm cm}^2
\ \ \ {\rm and}\ \ \ \tilde{l} \sim 10^{-6}\ {\rm cm}.
\label{HBRA6}
\end{equation} 
(iii) The hard prompt photon mean free path is then 
\begin{equation}
\frac{1}{L_\gamma}=R_{vac}\sigma_\gamma \approx 
4\alpha \left(\frac{\pi }{3}\right)^{1/3}
(\tilde{n}^{2/3}\tilde{l}) 
\label{HBRA7}
\end{equation}
in agreement with Eqs.(\ref{intro3}) and (\ref{intro4}). 
(iv) The final estimate for the mean free path of a hard prompt gamma 
photon follows from 
Eqs.(\ref{HBRA6}) and (\ref{HBRA7}) to be 
\begin{equation}
L_\gamma \sim 3.4\times 10^{-8}\ {\rm cm}.
\label{HBRA8}
\end{equation}
Thus, the hard photon is absorbed at a distance of less than a nanometer 
away from where it was first created. This constitutes the central result 
of this work.

\section{Conclusion \label{CON}}

Our picture of LENR in non-equilibrium situations near the surface of 
metallic hydrides may be described in the following manner: from the 
weakly coupled proton and electronic surface plasmon polariton  
oscillations, the electrons have their mass substantially renormalized 
upward. This allows for the production of ultra low momentum neutrons 
and neutrinos from heavy electrons interacting with protons or deuterons   
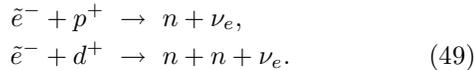
\begin{eqnarray}
\tilde{e}^-+p^+ &\to& n+\nu_e,
\nonumber \\ 
\tilde{e}^-+d^+ &\to& n+n+\nu_e.
\label{CON1}
\end{eqnarray}
The resulting ultra low momentum neutrons catalyze a variety of different 
nuclear reactions, creating complex nuclear reaction networks and related 
transmutations over time. The prompt hard 
gamma radiation which accompanies the neutron absorption is absorbed 
by the heavy electrons which drastically lowers the radiation frequencies 
of the finally produced photons via 
\begin{equation}
\tilde{e}_i ^- +\gamma _{hard}\to \tilde{e}_f ^- +\sum \gamma_{soft}.
\label{CON2}
\end{equation}
In the range of energies \begin{math}  \hbar \omega_\gamma  \end{math}
less than the renormalized energy of the heavy electrons, the prompt photon 
in Eq.(\ref{CON2}) impies a prompt hard gamma mean free path of less 
than a nanometer. The metallic hydride surface is thus opaque to hard photons 
but not to softer X-ray radiation in the KeV regime. In certain non-equilibrium 
metallic hydride systems, surface heavy electrons play 
a dual role in allowing both Eqs,(\ref{CON1}) for catalyzing LENR and 
Eq.(\ref{CON2}) for absorbing the resulting hard prompt photons.
Thus, the heavy surface electrons can act as a gamma ray shield.
Once the non-equilibrium conditions creating heavy electrons cease, 
ultra low momentum neutron production and gamma absorption both stop 
very rapidly.

\end{document}